\documentclass{article}
\usepackage[english]{babel}
\usepackage{float}
\usepackage{booktabs}
\usepackage{graphicx}% Include figure files
\usepackage{dcolumn}% Align table columns on decimal point
\usepackage{bm}% bold math
\usepackage{cite}
\usepackage{amsthm} % the proof of mathematical formula
\include{cip-v3-pack-and-defined}

\usepackage[letterpaper,top=2cm,bottom=2cm,left=3cm,right=3cm,marginparwidth=1.75cm]{geometry}

\usepackage{amsmath}
\usepackage{graphicx}
\usepackage[colorlinks=true, allcolors=blue]{hyperref}
\usepackage{authblk}

\title{Beyond Pairwise Interactions: Unveiling the Role of Higher-Order Interactions via Stepwise Reduction}
\author[a,1]{Junhao Bian}
\author[a,1]{Tao Zhou}
\author[a,1,$\ast$]{Yilin Bi}
\affil[a]{CompleX Lab, School of Computer Science and Engineering, University of Electronic Science and Technology of China, Chengdu 610054, China}
\affil[1]{Junhao Bian, Tao Zhou and Yilin Bi contributed equally to this work.}
\affil[$\ast$]{To whom correspondence should be addressed: \href{betayilin@gmail.com}{betayilin@gmail.com}}

\begin{document}
\maketitle

\begin{abstract}
Complex systems, such as economic, social, biological, and ecological systems, usually feature interactions not only between pairwise entities but also among three or more entities. These multi-entity interactions are known as higher-order interactions. Hypergraph, as a mathematical tool, can effectively characterize higher-order interactions, where nodes denote entities and hyperedges represent interactions among multiple entities. Meanwhile, all higher-order interactions can also be projected into a number of lower-order interactions or even some pairwise interactions. Whether it is necessary to consider all higher-order interactions, and whether it is with little loss to replace them by lower-order or even pairwise interactions, remain a controversial issue. If the role of higher-order interactions is insignificant, the complexity of computation and the difficulty of analysis can be drastically reduced by projecting higher-order interactions into lower-order or pairwise interactions. We use link prediction, a fundamental problem in network science, as the entry point. Specifically, we evaluate the impact of higher-order interactions on link predictive accuracy to explore the necessity of these structures. We propose a method to decompose the higher-order structures in a stepwise way, thereby allowing to systematically explore the impacts of structures at different orders on link prediction. The results indicate that in some networks, incorporating higher-order interactions significantly enhances the accuracy of link prediction, while in others, the effect is insignificant. Therefore, we think that the role of higher-order interactions varies in different types of networks. Overall, since the improvement in predictive accuracy provided by higher-order interactions is significant in some networks, we believe that the study of higher-order interactions is both necessary and valuable.

\end{abstract}


\section{Introduction}
From individual interactions in social relationships, to species symbiosis in ecosystems, to stock market fluctuations in economic markets and information flow through the Internet, real-world complex systems exhibit diverse behavioral patterns and time-varying characteristics via multi-level and multi-scale interactions, offering insights into natural and social phenomena \cite{sterman1994, arthur1999,ladyman2013}. The complexity of such a real system is not entirely attributable to the complication of system operational rules, but largely originated from the interactions between entities in the system \cite{kauffman1996,hidalgo2018}. Networks, as a powerful mathematical tool, have been naturally employed to depict and analyze the interactions between entities in complex systems \cite{posfai2016,newman2018}. In traditional networks, nodes denote entities of the system and links denote pairwise interactions between two nodes. However, interactions in real-world systems are not always simply pairwise relationships, while the collective interactions between multiple entities are pervasive \cite{battiston2021, battiston2020, battiston2022, bianconi2021, bick2023}. For example, in the scientific collaboration, if a project involves multiple researchers, pairwise interactions are insufficient, while higher-order interactions provide a more suitable framework to capture the feature of such multilateral collaboration \cite{bonacich2004}. In commercial transactions, third-party agents such as intermediaries may be involved in addition to buyers and sellers, and higher-order interactions involving multiple stakeholders more accurately reflect the business relationships \cite{bonacich2004}. In ecosystems, interactions between two microbial species are often regulated by other species \cite{wootton2002, werner2003, poisot2015, bairey2016}. For example, species A produces an antibiotic to inhibit species B, while a third species, C, secretes an enzyme that degrades this antibiotic, thereby reducing the inhibitory effect of species A on species B. This pattern of interactions among three microbial species cannot be simply captured by pairwise interactions, as species C introduces an additional regulatory layer that influences the interaction between species A and B \cite{bairey2016}. Similar scenarios involving higher-order interactions are numerous, such as biochemical reactions that often involve multiple substances or the intervention of enzymes \cite{klamt2009}, and proteins form complexes through interactions among multiple molecules \cite{gavin2002}.

The importance of higher-order interactions has already been noted by many scientists, for example, higher-order interactions can help us more accurately predict which drugs may cause adverse side effects when taken together (these effects don’t arise when the drugs are used alone) \cite{nguyen2023, vaida2019}, higher-order interactions are crucial to understand the dependencies among groups of neurons in avalanche dynamics \cite{yu2011}, and higher-order interactions can be employed to improve the accuracy of predictions in the visual responses of neurons in anesthetized cats \cite{ganmor2011}. At the same time, Zhang \textit{et al.} \cite{zhang2018} and Benson \textit{et al.} \cite{benson2018} have pointed out that the variability in the number of nodes contained in higher-order edges can pose both complexity and sophistication challenges to the network analysis, and Zhang and Chen \cite{chen2018} further found that graph neural networks (GNNs) highly effective in processing pairwise interactions but struggle to directly handle higher-order interactions. Considering the additional complexity involved in analyzing higher-order interactions, it remains to be determined whether the introduction of these interactions is a significant innovation or merely a costly topological game \cite{wolf2016, torres2021}.

In this study, we select link prediction, a fundamental task in network science, as an entry point to quantitatively analyze the effects of higher-order interactions. Link prediction aims to predict the links that exist but have not yet been observed, or that will appear in the future based on the observed network structure \cite{lu2011, martinez2016, zhou2021}. Hyperedge prediction is a natural extension of link prediction to hypergraphs, with prediction object being extended from pairwise links to higher-order edges \cite{chen2023}, which has become an active branch in the hypergraph study, with various prediction algorithms proposed \cite{xu2013, zhang2018, tu2018, zhang2019, kumar2020, hwang2022, contisciani2022}. However, the mainstream study has yet to consider the impacts of structures of different orders on the accuracy of prediction, nor has it analyzed the necessity of higher-order edges in the prediction of lower-order edges. Recently, Yoon \textit{et al.} \cite{yoon2020} proposed the concept of $n$-projected graphs, where each node represents a set of $n-1$ nodes, and an edge is formed between two such sets if their union of nodes contains exactly $n$ nodes and belongs to at least one hyperedge. They argued that an aggregation from 2-projected to $n$-projected graphs is a natural extension of the pairwise projected graph of a hypergraph (named as “collaboration networks” in earlier literature \cite{Newman2001}). Their experimental results demonstrated that augmenting a pairwise projected graph with a 3-projected graph can notably improve the prediction accuracy, with diminishing returns at higher orders. In addition, if one considers very high order projection, the prediction accuracy may decline. Although Yoon \textit{et al.}’s work provides a valuable starting point for understanding the role of higher-order interactions in link prediction, their projection-based approach results in an aggregated graph with a huge number of new-type nodes, each representing a collection of several original nodes, which largely complicates computation and analysis. Moreover, their method does not cleanly strip away the effects of higher-order interactions in a stepwise way. For example, if we wish to keep the information up to 5-order interactions and thus analyze the effects of 6-order and $6^{+}$-order interactions, we can not achieve this by analyzing the aggregation from 2-projected to 5-projected graphs, because the hyperedges of 3 to 5 orders are also projected into a number of pairwise relations. Therefore, no matter how many orders are taken into account, the method of Yoon \textit{et al.} will project all higher-order interactions to pairwise interactions, except that the set of nodes is different. Inspired by this study, we propose a more applicable method, called $n$-reduced graph, which retains the structural information of lower-order hyperedges while rationally decomposing higher-order interactions, significantly reducing the computational cost. Using this method, we can quantitatively investigate the role of higher-order interactions and efficiently predict hyperedges.

\section{Methods}
\subsection{Preliminary}
A hypergraph is denoted as $G(V, E)$, where $V = \{v_1, v_2,\dots, v_N \}$ denotes the set of nodes, and $E = \{e_1, e_2,\dots, e_M \}$ denotes the set of hyperedges \cite{berge1984, bretto2013}. In contrast to traditional pairwise interaction networks, where a link represents a connection between only two nodes, a hyperedge captures the relationship among multiple nodes, meaning that a hyperedge $e_\alpha$ can involve two or more nodes. The order of a hyperedge $e_{\alpha}$, denoted by $k_{\alpha}$, is the number of nodes contained in this hyperedge. The number of hyperedges involving node $v_{i}$ is defined as the hyperdegree of $v_{i}$, denoted as $d_i$. Hyperedge prediction aims to predict unobserved hyperedges based on the observed structure. Specifically, this algorithm tries to identify whether any unobserved hyperedge $e_c \subseteq 2^{V} \setminus E$ in the candidate set is indeed a true hyperedge \cite{chen2023}.

\subsection{The $n$-reduced graph}
\begin{figure}[htbp]
\centering
\centerline{\includegraphics[width=1\linewidth]{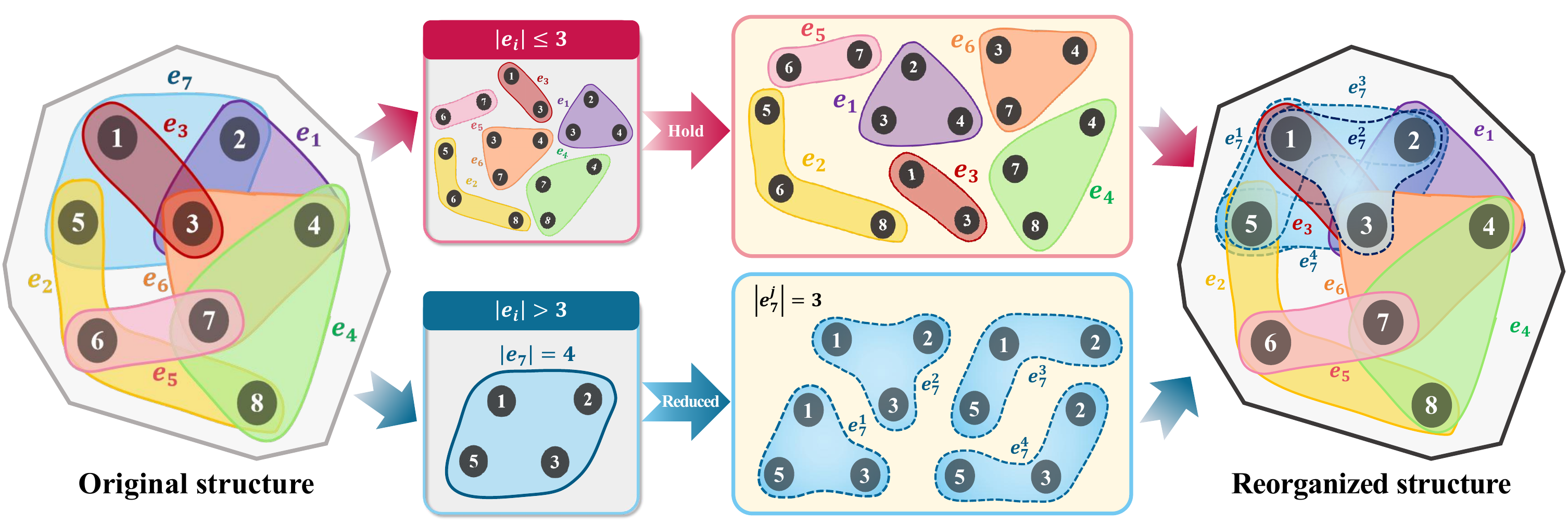}}
\caption{{\bf The process to construct the $n$-reduced graph.} This hypergraph contains 8 nodes and 7 hyperedges. Among them, the orders of $e_{2}, e_{3}$ and $e_{5}$ are 2, the orders of $e_{1}$, $e_{4}$ and $e_{6}$ are 3, and the order of $e_{7}$ is 4. When $n=3$, hyperedge $e_{7}$ is decomposed into all combinations of its nodes, say \{1,2,3\}, \{1,2,5\}, \{1,3,5\} and \{2,3,5\}.}
\label{fig_Process}
\end{figure}

In order to analyze the role of all hyperedges with orders larger than a threshold order, we propose the following method, referred to as the $n$-reduced graph. Given a hypergraph $G(V, E)$, its corresponding $n$-reduced graph, denoted by $G_n(V, E_n)$, is defined as
\begin{equation}
    E_n := \{e_\alpha: e_\alpha \in E, |e_\alpha| \leq n\} \cup \{e: e_\beta \in E, |e_\beta| > n, e \subseteq e_\beta, |e| = n \}.
\end{equation}
As defined above, the hyperedge set $E_n$ comprises two parts: (1) all hyperedges in $G$ with $k_\alpha \le n$ are retained in $E_n$; (2) the hyperedges with $k_\alpha > n$ are decomposed into $\binom{k_\alpha}{n}$ hyperedges, each of which has an order of $n$. Note that when $n=2$, $G_n$ is equivalant to the pairwise projection of $G$; and when $n=\underset{\alpha}{max} (k_\alpha)$, $G_n$ is equivalent to the original hypergraph $G$. Unlike the $n$-projected graph, the $n$-reduced graph retains the node set and all hyperedges with order no greater than $n$, while decomposing hyperedges of order greater than $n$ using sets of hyperedges with order $n$. In summary, the $n$-reduced graph is an approximated representation of the original hypergraph that consists of hyperedges of orders no greater than $n$, with the goal of minimizing information loss. It is particularly suitable for evaluating the role of hyperedges with orders larger than a threshold. Figure.~\ref{fig_Process} illustrates the construction process of a $3$-reduced graph. This hypergraph consists of 8 nodes and 7 hyperedges, with the orders of six hyperedges ($e_1, e_2, e_3, e_4, e_5, \text{and} \ e_6$) not exceeding 3. These hyperedges are directly retained in the $3$-reduced graph. For the hyperedge $e_7$ with $k_\alpha > 3$, we generate all 3-order hyperedges by traversing all triples from the set of nodes in the hyperedge $e_7$, which are subsequently added to the $3$-reduced graph.

\subsection{Features and Classifier}
This study applies three local similarity features and three weighted features to evaluate the likelihood of a set of nodes to form a hyperedge. The former measures whether the nodes in the set are all adjacent to some other nodes (two nodes are adjacent if they appear together in at least one hyperedge), while the latter measures the closeness between pairs of nodes. In brief, the former treats the candidate node set as a whole, whereas the latter views it as the sum of multiple pairwise relationships.

In this study, local similarity features are direct extensions of classic similarity indices used in link prediction \cite{barabasi1999, liben2007, zhou2009, newman2001, jaccard1901, adamic2003}. (1) \textbf{Common Neighbor Index}: The common neighbors of a hyperedge $e$ are the common neighbors of all nodes in $e$, that is
\begin{equation}
CN(e) = \left| \bigcap_{v \subseteq e} N(v) \right|,
\end{equation}
where $N(v)$ represents the neighbor set of the node $v$. (2) \textbf{Jaccard Coefficient Index}: The Jaccard coefficient is a normalization index that divides the number of common neighbors among all nodes in hyperedge $e$ by the total number of neighbors of any nodes in $e$:
\begin{equation}
JC(e) = \frac{\left| \bigcap_{v \subseteq e} N(v) \right|}{\left| \bigcup_{v \subseteq e} N(v) \right|}.
\end{equation}
(3) \textbf{Adamic Adar Index}: The Adamic Adar index penalizes high-degree nodes by dividing the degrees of common neighbors, as 
\begin{equation}
AA(e) = \sum_{v_i \in \bigcap_{v \subseteq e} N(v)} \frac{1}{\log d_i}.
\end{equation}

Weighted features measure the closeness among the candidates node set by using different averages of the weights of direct connections between node pairs. Let $W_{uv}$ denote the weight of the connection between nodes $u$ and $v$, which is defined as the number of hyperedges containing both $u$ and $v$. Clearly, if $u$ and $v$ are adjacent, then $W_{uv} > 0$, otherwise, $W_{uv} = 0$. For any candidate node set $e$, let $E_e$ represent the set of adjacent node pairs in $e$, that is
\begin{equation}
E_e:=\left\{(u,v): u\in e, v\in e, W_{uv}>0 \right\}.
\end{equation}
Note that $u \in e$ and $v \in e$ do not necessarily imply $W_{uv} > 0$, as $e$ may not be a hyperedge in the original hypergraph. Accordingly, we can calculate the average weight of all adjacent node pairs in $e$ using three different methods. (4) \textbf{Geometric Mean}:
\begin{equation}
GM(e) = \left( \prod_{(u,v) \in E_e} W_{uv} \right)^{\frac{1}{|E_e|}}.
\end{equation}
(5) \textbf{Harmonic Mean}: 
\begin{equation}
HM(e) = \frac{|E_e|}{\sum_{(u,v) \in E_e} W_{uv}^{-1}}.
\end{equation}
(6) \textbf{Arithmetic Mean}:
\begin{equation}
AM(e) = \frac{\sum_{(u,v) \in E_e} W_{uv}}{|E_e|}.
\end{equation}

This study integrates the six features into a hyperedge feature vector and utilizes the Logistic regression to train the model \cite{hosmer2013}. That is, the feature vector of a candidate node set $e$ can be represented as a linear combination of $[CN(e), JC(e), AA(e), GM(e), HM(e), AM(e)]$, and then this feature vector is used to learn the scoring function $f(e)$, such that:
\begin{equation}
    f(e) =
    \begin{cases}
    \geq \epsilon, & \text{if } e \in E \\
    < \epsilon, & \text{if } e \notin E
    \end{cases},
\end{equation}
where $\epsilon$ is the binary classification threshold of $f(e)$, used to determine whether $e$ is a potential hyperedge.

\subsection{Data Splitting and Sampling}\label{s2.4}
This study randomly splits each dataset into a training set and a test set in an 8:2 ratio. The training set is utilized to train the model, while the test set is used to evaluate the performance. For a given $n$, hyperedges in the training set with orders greater than $n$ are decomposed into multiple $n$-order hyperedges using the $n$-reduced operator. To enhance the model's generalization capability, 5-fold cross-validation is performed on the training set. Specifically, the training set is randomly divided into five equal-sized subsets. In each cross-validation iteration, one subset is designated as the validation set, while the remaining four subsets are used to calculate feature vectors, maintaining an 8:2 ratio between training and validation data. This process is repeated five times, ensuring that each subset is treated as the validation set once.

To address the significant disparity between the number of candidate hyperedges and true hyperedges in hyperedge prediction (note that, this disparity is much larger than that in link prediction for pairwise interaction networks, because the number of possible hyperedges in a hypergraph is about $2^N$, much larger than $N^2$ in a pairwise interaction network), this study adopts a negative sampling strategy, which generates non-existent hyperedges (negative samples) to balance with the missing hyperedges (positive samples) in both size and distribution \cite{patil2020, hwang2022}. For each positive hyperedge $e$ in the validation set or the test set, we randomly remove one node $v_0$ from $e$ and then randomly select a node from neighbors of $e$ to replace $v_0$. If the resulting set of nodes, $e_{neg}$, does not belong to the hyperedge set $E_n$, it is considered to be a valid negative sample. This negative sampling method ensures that the generated negative samples are structurally similar to positive hyperedges, thereby increasing the difficulty of prediction. Consequently, this method effectively enhances the model's generalization ability, enabling it to more accurately distinguish between existent and non-existent hyperedges.

\subsection{Evaluation Metrics}
Hyperedge prediction is a specialized binary classification task aimed at accurately distinguishing between positive and negative samples. To evaluate the predictive performance of the model, we utilize the Area Under the ROC Curve (AUC) as the evaluation metric \cite{fawcett2006}. AUC is widely used for evaluating classification models, and our previous works \cite{zhou2023, jiao2024, wbjz2024, Bi2024} demonstrate that AUC provides superior discriminability and greater information content compared to other commonly used metrics. The range of AUC is $[0,1]$. If the prediction is assigned completely at random, the AUC will be 0.5. A higher AUC indicates better predictive performance. Although some studies suggest that AUC may exhibit evaluation bias due to imbalanced positive and negative samples \cite{lobo2008, yang2015,carclo2024}, the negative sampling method employed in this study ensures a balanced number of positive and negative samples, thereby mitigating these issues. To enhance the robustness of the results, we conducted 10 independent experiments and used the average AUC as the final outcome.

\begin{table}[h]
\centering
\caption{{\bf The Statistics of the Dataset.} Here, $N$ is the number of nodes, $M$ is the number of hyperedges, and $\left \langle k \right \rangle$ is the average order of hyperedges.}
\label{datasets}
\begin{tabular}{p{4.5cm} p{1.8cm} p{1.8cm} p{1.8cm} p{1.8cm}}
\toprule
\textbf{Dataset} & \textbf{$N$} & \textbf{$M$} & \textbf{$\left \langle k \right \rangle$} & \textbf{Reference} \\
\midrule
email-Enron      & 143  & 1466  & 2.95   & \cite{benson2018}  \\ 
email-Eu         & 530  & 1119  & 2.51   & \cite{benson2018, leskovec2007, yin2017}  \\
threads-math-sx  & 2563 & 2918  & 2.50   & \cite{benson2018}  \\
DAWN             & 895  & 7728  & 3.15   & \cite{benson2018}  \\
NDC-classes      & 1140 & 940   & 4.43   & \cite{benson2018}  \\
NDC-substances   & 2647 & 4606  & 4.87   & \cite{benson2018}  \\
\bottomrule
\end{tabular}
\end{table}

\section{Data}
This study utilizes six real-world hypergraphs of different sizes. \textbf{email-Enron:} This data originates from the internal email communications of Enron Corporation, where nodes represent employees' email addresses, and a hyperedge comprises the sender and all recipients of an email. \textbf{email-Eu:} This data comes from the internal email communications of a European research institution, where nodes represent the members' email addresses, and the sender and all recipients involved in an email form a hyperedge. \textbf{threads-math-sx:} This data is sourced from \textit{math.stackexchange.com}, where nodes represent users on the website, and a hyperedge consists of all users who participated in a discussion thread within 24 hours. \textbf{DAWN:} This data originates from the Drug Abuse Warning Network (DAWN) in the United States, recording nationwide emergency department visits related to drug use, where nodes are unique drug IDs, and hyperedges comprise all drugs used in the same emergency incident. \textbf{NDC-classes:} This data is based on the National Drug Code (NDC) directory of the United States, where nodes denote drug class labels, and each hyperedge represents multiple class labels involved in one drug. \textbf{NDC-substances:} In this data, nodes represent individual drugs, and hyperedges are the substances composing these drugs. Since hyperedges containing at least 10 nodes are rare and the computational complexity involving these structures is huge, this study, like \cite{yoon2020}, only considers hyperedges with $k \le 10$. Table~\ref{datasets} shows basic statistics of these six real-world networks, including the number of nodes and hyperedges, and the average order of hyperedges.

\begin{figure}[htbp]
\centering
\centerline{\includegraphics[width=0.95\linewidth]{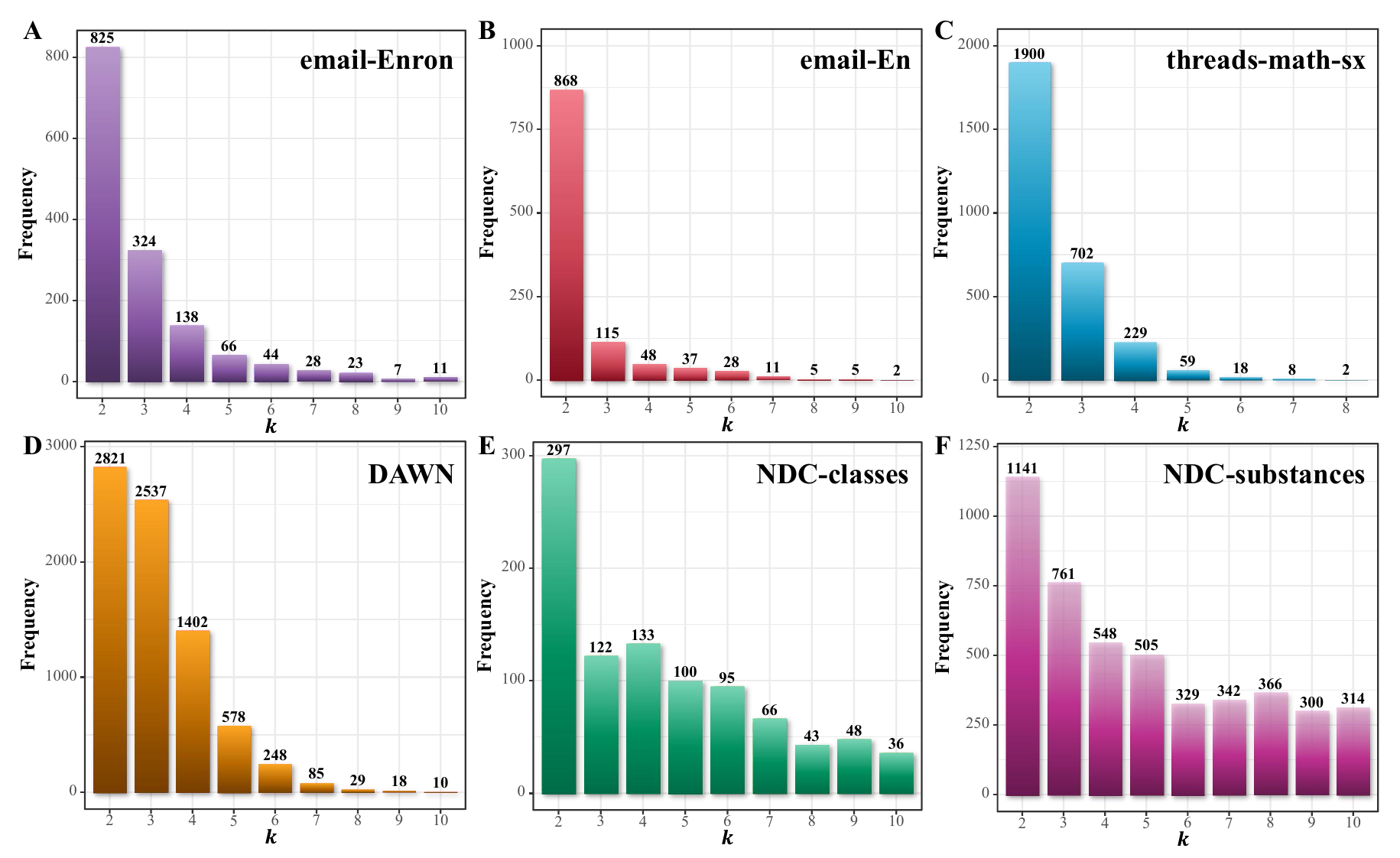}}
\caption{{\bf The frequency distribution of hyperedge orders.} The x-axis represents the order $k$ of hyperedges, while the y-axis indicates the number of hyperedges containing $k$ nodes.}
\label{HyperEdge_D}
\end{figure}

\begin{figure}[htbp]
\centering
\centerline{\includegraphics[width=1.05\linewidth]{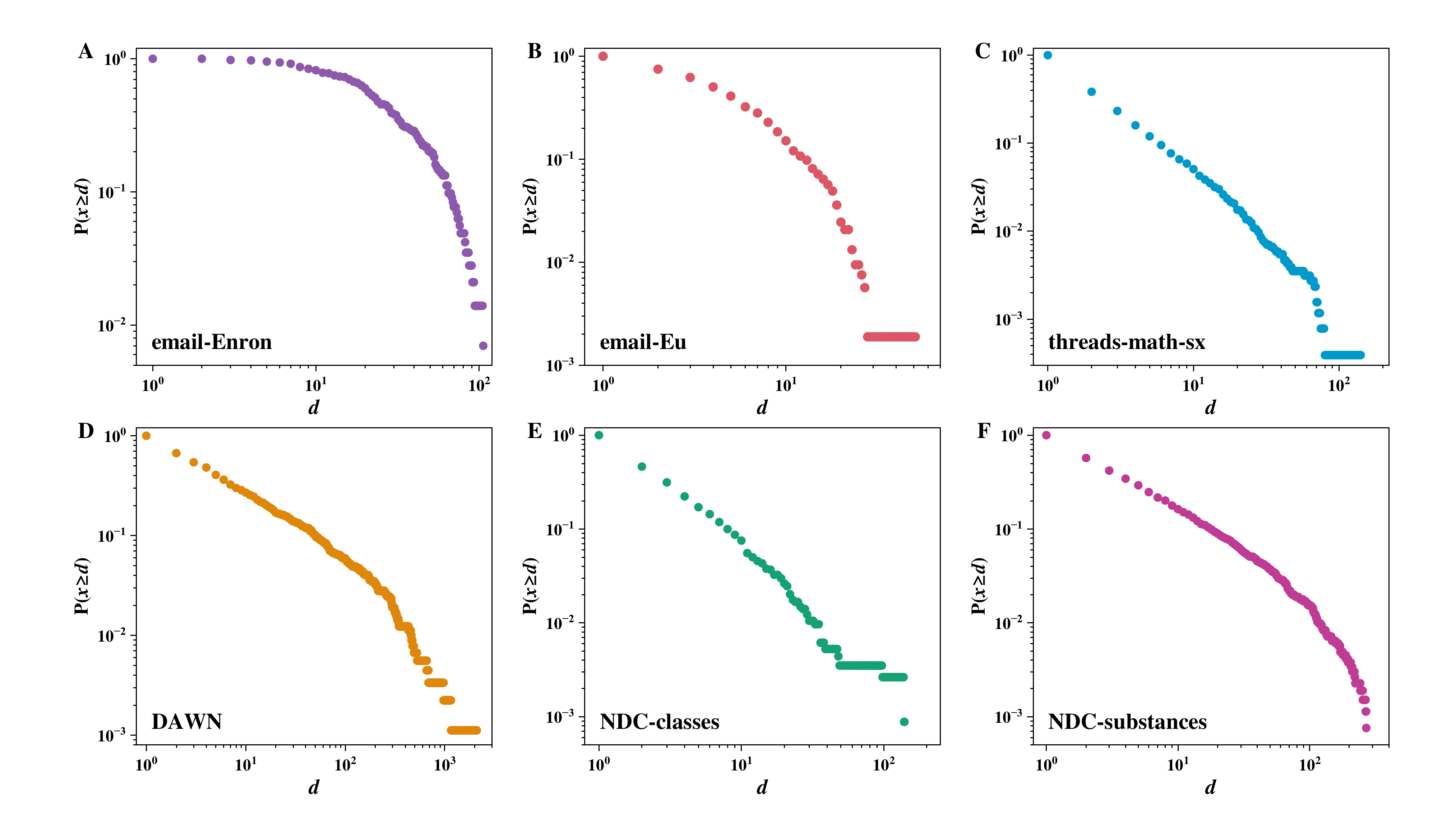}}
\caption{{\bf The cumulative distribution of hyperdegrees.} The x-axis represents the hyperdegree $d$, while the y-axis represents the cumulative probability distribution $P(x \geq d)$. Each distribution is exhibited in a log-log scale.}
\label{Hyperdegree}
\end{figure}

Figure.~\ref{HyperEdge_D} illustrates the distribution of hyperedge orders. It is evident that lower-order hyperedges constitute the majority, while the number of hyperedges decreases significantly as the order increases. Figure.~\ref{Hyperdegree} shows the cumulative distribution of hyperdegrees. These distributions exhibit a broad range and display similar characteristics to power-law distributions, though they cannot be precisely characterized by power laws \cite{holme2019, broido2019}.

\section{Results}
\begin{figure}[htbp]
\centering
\centerline{\includegraphics[width=1\linewidth]{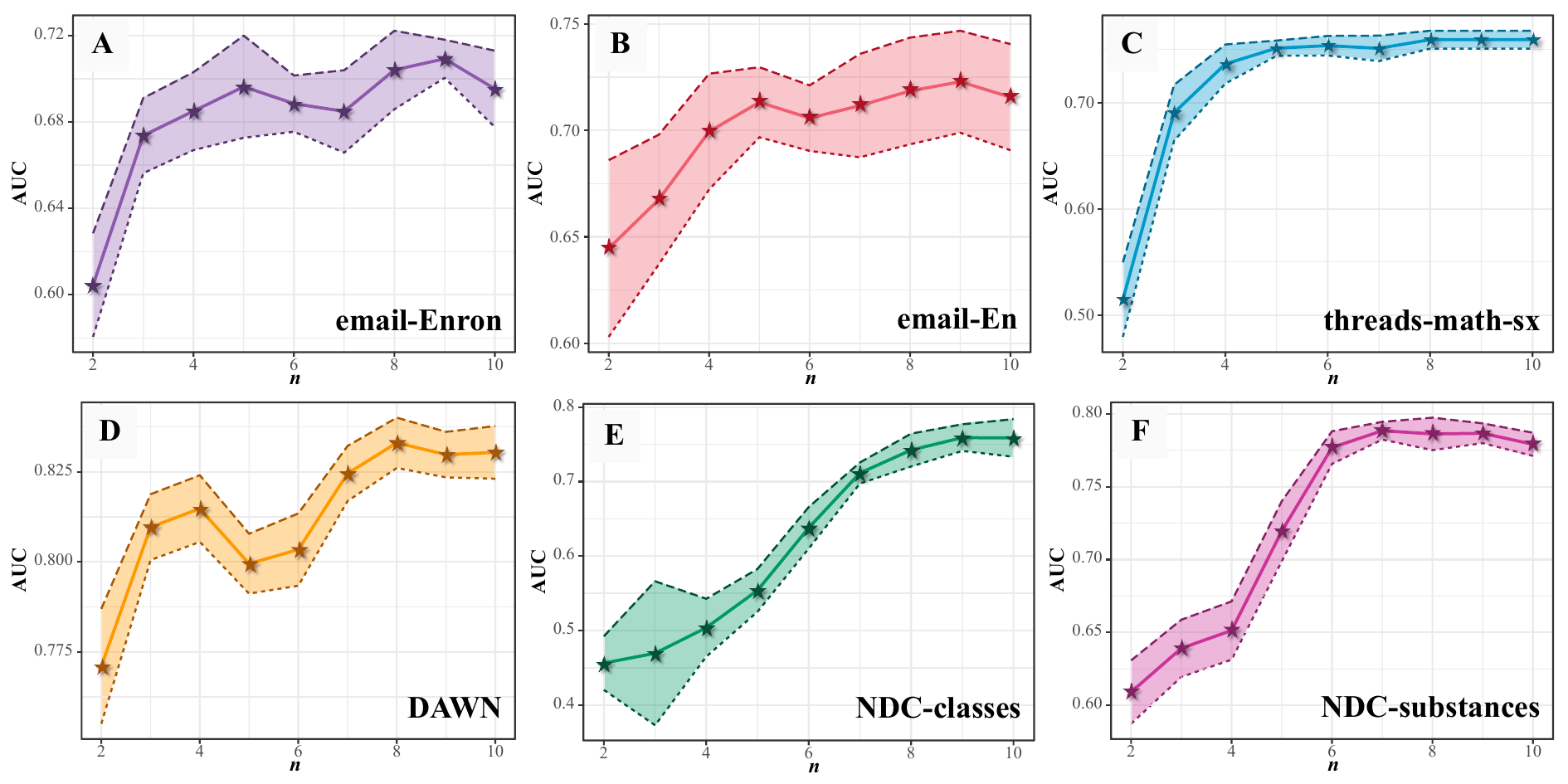}}
\caption{{\bf The change of the average AUC with increasing $n$ in the $n$-reduced graph.} The solid line represents the average AUC over 10 independent experiments, and the shaded area represents the range of the average AUC plus and minus the standard deviation.}
\label{nr1}
\end{figure}

In a given hypergraph, all hyperedges of orders $k > n$ are decomposed into multiple $n$-order hyperedges by the $n$-reduced operator. Clearly, the larger of $n$, the more higher-order information is retained. As shown in Section.~\ref{s2.4}, to ensure consistency in evaluating the training and testing of hyperedge structures, hyperedges of order $k > n$ are excluded from the test set, allowing predictions to focus solely on hyperedges with $k \leq n$. Figure.~\ref{nr1} illustrates the trend of AUC with increasing $n$ across six hypergraphs. In all hypergraphs, AUC generally increases with $n$, indicating that higher-order information enhances prediction accuracy. However, as $n$ increases, the change in AUC tends to level off, suggesting that the incremental benefit from additional higher-order information diminishes for large $n$.

\begin{figure}[htbp]
\centering
\centerline{\includegraphics[width=0.95\linewidth]{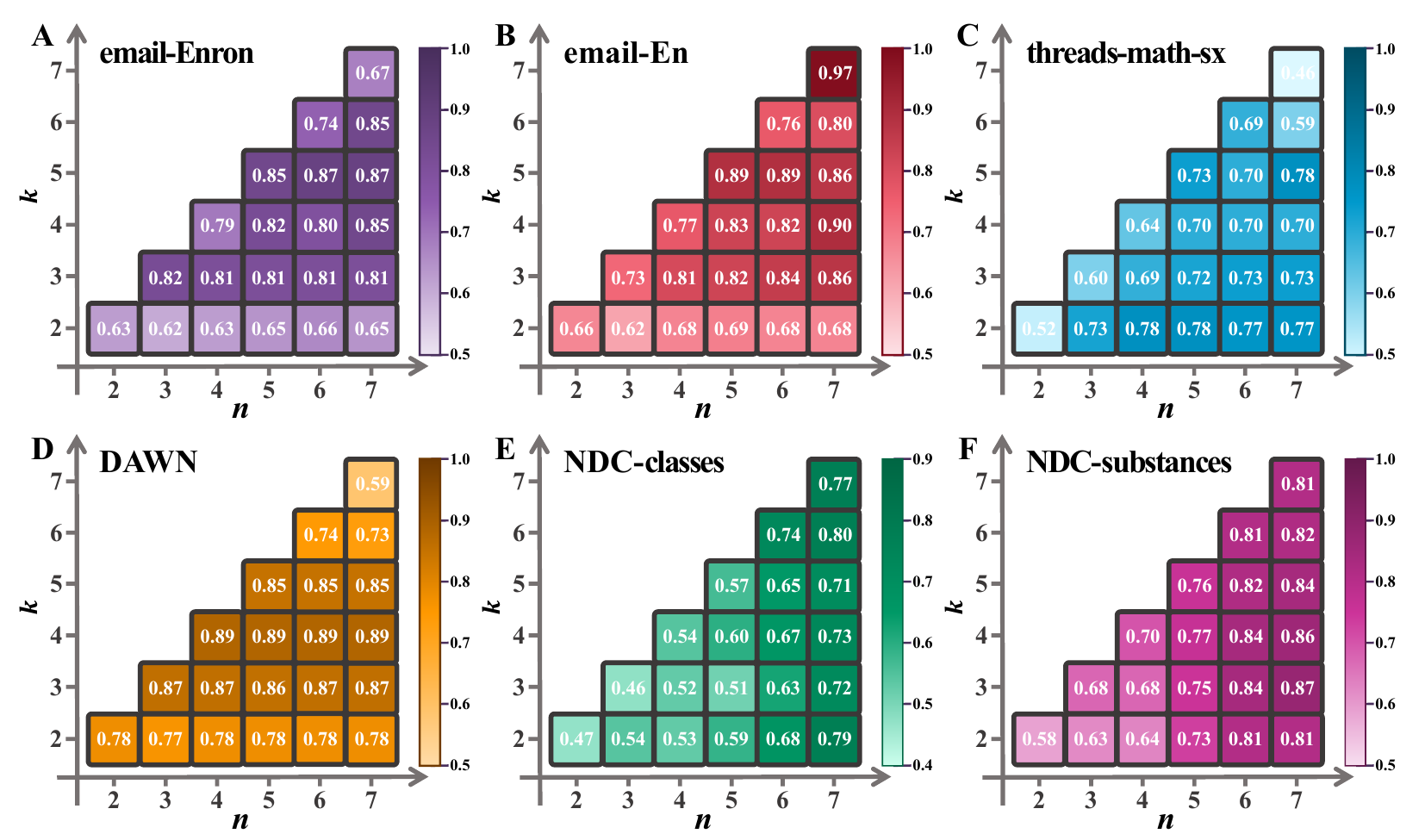}}
\caption{{\bf The values of AUC in predicting solely $k$-order hyperedges in the $n$-reduced graph.} The shade of color indicates the value of AUC, with darker colors corresponding to larger AUC values.}
\label{nr2}
\end{figure}

As shown in Figure.~\ref{nr1}, the average AUC shows an upward trend for every hypergraph, suggesting that higher-order interaction information significantly enhances the accuracy of hyperedge prediction. However, directly using these results to quantify the contribution of higher-order interactions to the predictive accuracy is not rigorous. This is because hyperedges in the test set vary with different $n$, making the reasons underlying different AUC values very complicated. To eliminate confusion caused by varying test sets, we further consider the prediction of hyperedges with a fixed order $k$ within the test set. For example, we can compare the predictive accuracies for hyperedges of order $k = 3$ in the test set across different $n$-reduced graphs ($n \ge 3$). This ensures the consistency of the test set and clearly highlights the impact of higher-order information. Figure.~\ref{nr2} illustrates the trend of AUC values as $n$ increases for different $k$. For $k \ge 3$, most networks (excluding email-Enron and DAWN) show a significant increase in AUC with the presence of more and more higher-order information. Particularly in the networks of NDC-classes and NDC-substances, the enhancement in prediction accuracy with higher-order information is most evident. For example, when $k = 3$, as $n$ increases from 3 to 7, the AUC rises from 0.46 to 0.72 and from 0.68 to 0.87, respectively. In these cases, decomposing higher-order hyperedges into lower-order ones, even if the resulting hyperedges are still of order no less than 3, will reduce the predictive accuracy for 3-order hyperedges, indicating that higher-order information is crucial for hyperedge prediction. For $k = 2$, except in the networks of NDC-classes and NDC-substances, where higher-order interactions notably improve the predictive accuracy of 2-order hyperedges, the increasing of AUC is relatively flat or even fluctuates in other networks with the addition of higher-order information. Additionally, some networks exhibit diminishing returns from higher-order information on prediction performance. As shown in Figures.~\ref{nr2}(A), \ref{nr2}(C), and \ref{nr2}(D), the increase in AUC tends to be flat with the increasing $n$, suggesting that reaching a certain level, the benefits of higher-order information diminish. Conversely, in Figures.~\ref{nr2}(E) and~\ref{nr2}(F), the AUC continues to rise significantly with increasing $n$, indicating that retaining more higher-order information can significantly enhance prediction performance in these networks. In a word, higher-order interactions are generally valuable, but their roles vary across different networks. In some networks, they play a vital role, while in others, their contributions are negligible.

\begin{figure}[htbp]
\centering
\centerline{\includegraphics[width=0.85\linewidth]{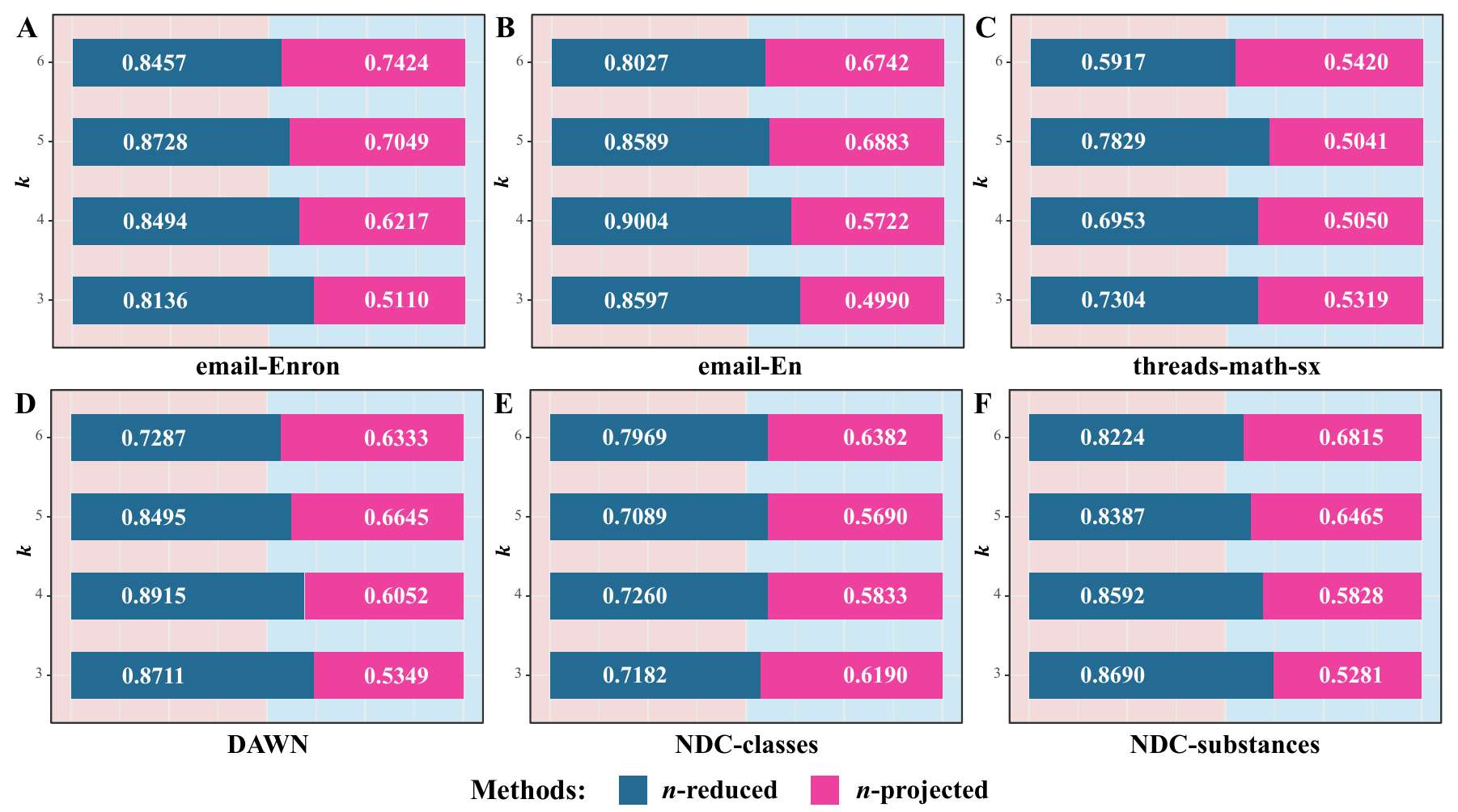}}
\caption{{\bf Comparison of AUC between $n$-reduced and $n$-projected operators for predicting $k$-order hyperedges.} Here, $n$ is set to 7, and the reported results are all averaged over 10 independent experiments. The y-axis represents the order of the predicted hyperedge. The dark blue bars denote the average AUC for the $n$-reduced method, while the magenta bars represent the average AUC for the $n$-projected method. To provide a clear comparison of prediction accuracy between the two methods, the length of each bar is scaled according to the AUC of each method as a proportion of the sum of AUC values of both methods. For instance, if the AUC for the $n$-reduced method is 0.8 and the AUC for the $n$-projected method is 0.7, the lengths of the dark blue and magenta bars would be  $8/15$ and $7/15$, respectively.}
\label{np&nr}
\end{figure}

In the Introduction, we highlight a method known as $n$-projected graph \cite{yoon2020}, which is also applicable for analyzing the impact of higher-order interactions on hyperedge prediction. Given a hypergraph $G(V, E)$, its $n$-projected graph is denoted as $G_ {n}^p=(V_{n}^p, E_{n}^p)$, where the node set $V_n^p$ is defined as the set of all subsets of $V$ consisting of $n-1$ nodes, with each subset represented by a single node, that is
\begin{equation}
V_n^p:=\{v_n \subseteq V : |v_{n}|=n-1\}.
\end{equation}
In $V_n^p$, node $v_n$ and node $u_n$ are linked if they satisfy the following two conditions: (1) $v_n$ and $u_n$ differ by exactly one element among the $n-1$ elements each contains, i.e., $|u_{n} \cup v_{n}|=n$; (2) the union of $v_n$ and $u_n$ is contained by at least one hyperedge in $E$. Accordingly,
\begin{equation}
E_n^p:=\{ (u_{n},v_{n}) \in V_{n}^{2} : |u_{n} \cup v_{n}|=n \enspace \& \enspace \exists e \in E, s.t., u_{n} \cup v_{n} \subseteq e \}.
\end{equation}
For a given $n$, the graph $G^p(n)$ used for further hyperedge prediction is constructed as a direct aggregation from the $2$-projected graph up to the $n$-projected graph, denoted as $G^p(n):=\left( G_2^p, G_3^p, \cdots, G_n^p \right)$. Notably, for any $n \geq 2$, $G^p(n)$ contains only pairwise interactions and its complexity primarily arises from the quantity and heterogeneity of nodes. Since $n$-projected and $n$-reduced graphs are two approaches that, while conceptually similar, decompose higher-order interactions in fundamentally different ways, we are particularly interested in which method retains more useful information. A natural assumption is that the one retains more useful information will achieve higher accuracy in hyperedge prediction. For comparison, we use the same six features and logistic regression model to predict hyperedges in the $n$-projected graph, following the method proposed by Yoon \textit{et al.} \cite{yoon2020}. Based on an edge's attribute in $G^{p}(n)$, we can identify its order in the original hypergraph. For instance, an edge between two sets of 4 nodes corresponds to a 5-order hyperedge in the original hypergraph. Figure.~\ref{np&nr} compares the prediction performance of these two methods for the hyperedge orders ranging from 3 to 6 with a fixed $n$, say $n=7$. The results show that the $n$-reduced method significantly outperforms the $n$-projected method across all six real networks and different orders. Specifically, the best prediction performance of the $n$-projected graph is $\rm AUC=0.74$, with an average performance as $\rm AUC=0.60$, whereas the $n$-reduced graph achieves a best prediction of $\rm AUC=0.90$ and an average prediction performance as $\rm AUC=0.81$. These findings suggest that if higher-order interactions need to be reduced for convenient and efficient analysis or some other reasons, the method of the $n$-reduced graph is superior in retaining useful higher-order information.

\section{Discussion}
Bipartite graphs \cite{Lambiotte2005, Shang2010} or collaboration networks \cite{Newman2001, Zhang2006} were usually used to represent higher-order interactions before hypergraphs became a popular tool in network science. Using scientific collaborations as an example, hypergraphs can effectively represent the collaboration of scientists in publishing articles, where nodes represent scientists. When multiple scientists coauthorize an article, they collectively form a hyperedge. For instance, if an article has five authors, this hyperedge will consist of five nodes. In contrast, when using bipartite graphs, nodes typically have two different types: one represents authors, while the other represents articles, with each article linked to all its authors. Alternatively, if we use collaboration networks to illustrate scientific collaboration, each node represents a scientist, and a link is drawn between two scientists if they are co-authors on at least one article. Mathematically, bipartite graphs and hypergraphs are equivalent, allowing bipartite graphs to represent higher-order interactions without losing any information. But why haven't bipartite graphs become the mainstream tool for analyzing higher-order interactions like hypergraphs? One reason is that bipartite graphs contain nodes with two different types, and this heterogeneity among nodes will increase the complexity of the analysis. Collaboration networks can be viewed as projections of higher-order interactions onto pairwise interactions. For example, a hyperedge with five authors can be projected into ten pairwise interactions among these authors. Although one can introduce edge weights to best retain edge information of higher-order interactions, this projection clearly leads to information loss compared to the lossless representation of bipartite graphs \cite{Newman2001b, Zhou2007, Wang2024}. Nevertheless, collaboration networks are still of great importance because they are easier to analyze than hypergraphs. An important question is: In which scenarios do all of the information from higher-order interactions need to be retained, and in which cases can they be represented as pairwise interaction networks such as collaboration networks? If they cannot be represented as pairwise interaction networks, can they be represented as lower-order hypergraphs to reduce the complexity of the analysis?

The key to addressing these questions is to determine the quantifiable impact of higher-order interactions in a given scenario. For example, in a network where tasks such as link prediction, community detection, critical nodes identification, and spreading prevalence estimation are considered, how can we ascertain whether it is necessary to consider hyperedges of 5 and $5^{+}$ orders? A simple approach is to remove all hyperedges with $k>4$ and compare the performance on specific tasks before and after the removal. However, this approach is too crude, because even if we apply the pairwise projection, some information about 5-order and $5^{+}$-order interactions will be retained. That is to say, if we cannot analyze these higher-order hyperedges because of the limited computational resources, we can at least project them into multiple pairwise interactions rather than remove them entirely. A less aggressive method is to directly compare hypergraphs with their pairwise projections \cite{Iacopini2019}. However, this results in a significant loss of information, and we cannot focus solely on the impacts of hyperedges of orders $k \ge 5$, since 3-order and 4-order interactions are also projected onto pairwise interactions. The optimal approach is to employ stepwise decomposition to retain as much information as possible. The effect of any information that has to be lost should then be attributed to the influence of higher-order interactions. For example, to analyze the effect of hyperedges of 5 and $5^{+}$ orders, we should strive to retain their information through interactions of 2-order to 4-order as much as possible before performing a comparative analysis.

The above-mentioned idea is used in both $n$-projected and $n$-reduced graphs to decompose higher-order interactions. However, they are different. The $n$-projected operator ultimately projects all higher-order interactions into pairwise interactions, introducing a large number of heterogeneous nodes to retain information about the higher-order edges as much as possible. In contrast, the $n$-reduced operator represents higher-order interactions through lower-order interactions while keeping the set of nodes unchanged. Although these two methods share the same starting point, the $n$-reduced graph may be more suitable to be an analyzing tool, because researchers seem to be less inclined to deal with heterogeneous nodes. Otherwise, bipartite graphs would be more popular than hypergraphs for representing higher-order interactions.

In this study, we compare the $n$-projected and $n$-reduced methods by using link prediction as an entry point to quantify the influence of higher-order interactions. The reason for the choice of link prediction is that it is a more fundamental task than the analysis of specific networked dynamics. From the experiment in this study, there are two main conclusions: First, we find that higher-order interactions generally have a significant and positive effect on predictive accuracy, but are not universally applicable to all networks. Second, compared to the $n$-projected method, the $n$-reduced method proposed in the study retains more higher-order information. This is evidenced by the fact that the $n$-reduced method achieves a higher accuracy in hyperedge prediction under the same conditions.

In conclusion, we offer three specific suggestions. First, our findings show that higher-order interactions have a significant and positive impact on hyperedge prediction. Therefore, it is recommended that researchers prioritize using hypergraphs to represent higher-order interactions when related analysis is feasible. Second, we have noticed that the influence of higher-order interactions varies significantly across different real-world networks. Therefore, in order to draw more comprehensive and reliable conclusions, one should perform further analysis on domain-specific and topology-specific hypergraphs. This may even include the distinction of which hyperedges bring richer information in a given hypergraph \cite{musciotto2021}. Finally, we find that the $n$-reduced operator effectively retains higher-order interaction information. Therefore, we recommend researchers further use the $n$-reduced graphs as an analytical tool to quantify the role of higher-order interactions on specific dynamics (e.g., propagation and synchronization) \cite{majhi2022, boccaletti2023, zhangy2023, Wang2024b} and other graph mining tasks (e.g., critical nodes identification and community detection) \cite{zeng2024,xiao2023,liu2023}.

% \section{Reference}
\bibliographystyle{plain}

\end{document}